\documentclass[prl,twocolumn,showpacs,final,superscriptaddress]{revtex4}
\usepackage{times}
\usepackage{bm}
\usepackage{graphics}
\usepackage{graphicx}
\usepackage{bm}
\usepackage{amsfonts}
\usepackage{amssymb}
\usepackage{amsmath}

\begin{document}
\title{Imaging of $s$ and $d$ partial-wave interference in quantum scattering of identical bosonic atoms}
\author{Nicholas R. Thomas}\affiliation{Department of Physics, University of Otago,
Dunedin, New Zealand} \author{Niels Kj\ae
rgaard}\email{nk@physics.otago.ac.nz}\affiliation{Department of
Physics, University of Otago, Dunedin, New Zealand}
\author{Paul S. Julienne}\affiliation{National Institute of
Standards and Technology, 100 Bureau Drive, Stop 8423,
Gaithersburg, Maryland, 20899-8423 USA} \author{Andrew C.
Wilson}\affiliation{Department of Physics, University of Otago,
Dunedin, New Zealand}

\date{\today}

\begin{abstract}
We report on the direct imaging of $s$ and $d$ partial-wave
interference in cold collisions of atoms. Two ultracold clouds of
$\rm ^{87}Rb$ atoms were accelerated by magnetic fields to collide
at energies near a $d$-wave shape resonance. The resulting halos
of scattered particles were imaged using laser absorption. By
scanning across the resonance we observed a marked evolution of
the scattering patterns due to the energy dependent phase shifts
for the interfering $s$ and $d$ waves. Since only two partial wave
states are involved in the collision process the scattering yield
and angular distributions have a simple interpretation in terms of
a theoretical model.
\end{abstract}
 \maketitle

\begin{figure}[]
\centerline{\includegraphics*[width=\columnwidth]{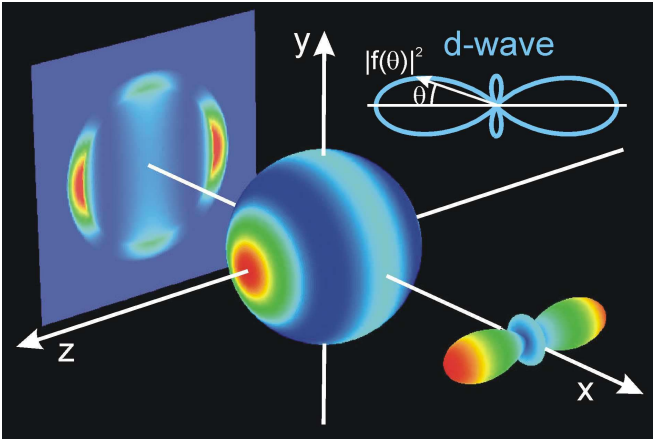}}
\caption{(Color online) Illustration of the process of using
absorption imaging for the detection of scattered particles. We
present the case of pure $d$-wave scattering occurring at the
origin for particles coming in along the $z$-axis. Scattered
particles will be situated on an expanding sphere and distributed
according the $d$-wave angular emission pattern $|f(\theta)|^2$.
Absorption imaging along the $x$-axis projects this distribution
onto the $yz$-plane.} \label{figratio}
\end{figure}
Collisions of atoms is a classic discipline of quantum mechanics
\cite{Faxen27,condon31}. With the advent of laser cooling
\cite{phillips98}, confining and cooling various atomic species to
submillikelvin temperatures became possible and gave rise to a
wealth of experiments in which quantum effects in collisions at
very low energies
--- cold collisions --- were observed \cite{Weiner99}.
Knowledge about cold collision properties paved the way for
exciting new developments in experimental atomic physics. It
played a crucial role in achieving Bose-Einstein condensation
\cite{anderson95} and Fermi degeneracy \cite{DeMarco1999a} in
dilute atomic vapors by mediating thermalization during
evaporative cooling and accounting for stability
\cite{Gerton2000a}. For atoms at temperatures associated with the
quantum degenerate regime the essential interaction properties are
determined by a single atomic parameter, the scattering length,
because all elastic scattering has an isotropic ($s$-wave) nature
at such low energies. The scattering length may exhibit a
pronounced dependency on external magnetic fields giving rise to
so-called Feshbach resonances \cite{Inouye1998a} which have
recently been exploited to create ultracold molecules and
molecular Bose-Einstein condensates (BECs)
\cite{Donley02,Regal03,Herbig03,Greiner03,Zwierlein04}.

To date, most experiments on cold collisions of atoms have been
carried out using magneto-optical traps (MOTs) or magnetic traps
which suffer from the disadvantages that no collision axis is
singled out or the collision energy cannot be varied over a wide
range \cite{Thorsheim90,Weiner99}. If no fixed collision axis is
present, anisotropic scattering, as occurs for collision energies
above the $s$-wave regime, will be obscured by spatial averaging.
One solution to this problem was provided in the ``juggling" MOT
experiment \cite{legere98}, where a cloud of cesium atoms was
laser cooled to 3 $\rm \mu K$ and ejected vertically from a trap
to collide with a previously launched cloud at energies up 160
$\mu K$. Scattered atoms were detected using a spectroscopic
technique revealing interference between $s$ and $p$ partial
waves. In experiments on BECs, a collision axis was also selected
using Bragg scattering to accelerate part of the atomic cloud, and
pure $s$-wave scattering halos were directly imaged
\cite{Chikkatur00}.

In this Letter, we report experiments on the collision of two
bosonic atomic clouds, initially confined in a magnetic
double-well potential and evaporatively cooled to a temperature
just above the phase transition for Bose-Einstein condensation.
Collisions at a selectable energy occur when the trapping
potential is continuously modified to a single-well configuration.
The atomic clouds accelerate from the sides of the harmonic
potential and collide at the center of the well. The resulting
scattering is equivalent to cold collisions of counter-propagating
ultracold pulsed atomic beams. Angular resolved detection of
scattered atoms is obtained using laser absorption imaging.
Specifically, we consider atomic clouds of doubly spin-polarized
$\rm ^{87}Rb$ which are cooled to a temperature of $\sim 225$~nK
and accelerated to collide with energies in the range from 87 to
553~$\rm \mu K$ as measured in units of Boltzmann's constant
$k_B$. In this energy interval a $d$-wave shape resonance is known
to occur \cite{Boesten1997a}. We observe scattering patterns
evolving from $s$-wave-like to $d$-wave-like distributions via
intermediate $s+d$ interfering scattering states which expose the
quantum mechanical origin of the process.

Quantum scattering of two particles under our conditions is
conveniently described in the partial wave formalism.  The wave
function for the relative motion is written $\psi =e^{ik
z}+f(\theta)e^{ikr}/r$, where $k$ is the magnitude of the relative
wave-vector of the colliding particles. The first term of this sum
represents an incoming plane wave travelling along the $z$-axis,
while the second term represents a radially outgoing scattered
wave with an amplitude which depends on the angle $\theta$ to the
$z$-axis, (see, e.g., \cite{Geltman69}). Using a partial wave
expansion of $\psi$, the scattering amplitude for identical bosons
is expressed as $f(\theta)=\sum_{l\,{\rm
even}}(2l+1)(e^{2i\eta_l}-1)P_l(\cos \theta)/ik$, where $P_l$ is
the Legendre polynomial of order $l$ and $\eta_l$ are the partial
wave phase shifts. The $l$th term in the expansion represents
particles having orbital angular momentum $l\hbar$ and the sum
only runs over even $l$, since odd partial waves are forbidden by
the requirement of a totally symmetric wave function for identical
bosonic particles. In the present case, where only $l=\{0,2\}$
terms ($s$ and $d$ waves) contribute \cite{note1} the scattering
amplitude is
\begin{equation}
f(\theta)=\{
\underbrace{(e^{2i\eta_0}-1)}_{s}+\underbrace{5(e^{2i\eta_2}-1)(3\cos^2\theta-1)/2}_{d}\}
/ik,
\end{equation}
and the differential cross section $d\sigma /d\Omega =
|f(\theta)|^2$ has an angular pattern which depends crucially on
the quantum mechanical interference between the partial wave
states as dictated by the phase shifts. Assuming the collisions to
occur in free space, scattered particles observed in the center of
mass frame will be distributed over a ballistically expanding
sphere (the so-called Newton sphere) according to the differential
cross section. If the scattered particles are detected using
absorption imaging, the distribution on this sphere will be
projected onto a plane by the Abel transformation
\cite{Bracewell00}. Figure~1 illustrates this in the case of pure
$d$-wave scattering and imaging along a direction perpendicular to
the collision axis.

In our experiment, $\rm ^{87}Rb$ atoms collected in a MOT were
optically pumped into the $\vert F = 2, M = 2 \rangle $ hyperfine
substate and loaded into a Ioffe-Pritchard magnetic trap in the
quadrupole-Ioffe-configuration \cite{Esslinger1998a}. The trapping
potential is cylindrically symmetric and harmonic, characterized
by radial and axial oscillation frequencies of
\mbox{$\omega_r/2\pi = 275$ Hz} and \mbox{$\omega_z/2\pi = 16$
Hz}, respectively. After radio-frequency evaporative cooling to a
temperature of \mbox{12 $\mu$K}, the trap was adiabatically
transformed to a double-well configuration \cite{Thomas02},
splitting the atomic cloud in half along it's long dimension ($z$)
by raising a potential barrier. The $z$-axis is horizontal so that
the influence from gravity is unimportant. An additional rotating
bias field of 0.5~mT is applied just before forming the
double-well to avoid Majorana spin-flip atom loss at the two trap
minima, where the magnetic field would otherwise become zero. When
fully separated the two clouds were 4.3~mm apart and the trap
frequencies were \mbox{$\omega_r/2\pi = 60$ Hz} and
\mbox{$\omega_z/2\pi = 14$ Hz} near the well minima. Further
evaporative cooling lowered the temperature to $~225$~nK in each
well (as compared to the BEC transition temperature of 100~nK),
and the total number of remaining atoms was \mbox{$ 4 \times
10^5$}. There is a slight difference between the properties of the
two clouds due to a small residual tilt in the potential.
Subsequently the separation of the clouds was adiabatically
adjusted to select the potential energy gained when the trap is
rapidly converted back to a single-well. To increase the cloud
densities the rotating bias field was reduced to 0.2~mT. The
collision is initiated by rapidly ramping from a double to a
single well configuration, accelerating the clouds towards the
potential minimum positioned between them. The trapping
configuration for the collision has frequencies
\mbox{$\omega_r/2\pi = $ 155 Hz} and \mbox{$\omega_z/2\pi = 12$
Hz}, and remains unchanged until the end of the experiment. After
the collision we waited for one quarter of the radial trap period,
so that atoms were at maximum radial extension, before acquiring
an absorption image \cite{ketterle1999a} by pulsing a resonant
laser beam onto the atoms. The 3D distribution of scattered atoms
is projected onto a plane giving the column density distribution.
We calculated the cloud velocities from positions measured before
and after the collision in additional experimental runs, and find
the collision energy, expressed in temperature units, using $T=\mu
v^2/2k_B$, where $\mu$ is the reduced mass of the particles and
$v$ is the relative velocity of the two clouds.

\begin{figure}[t!]
\centerline{\includegraphics*[width=0.98\columnwidth]{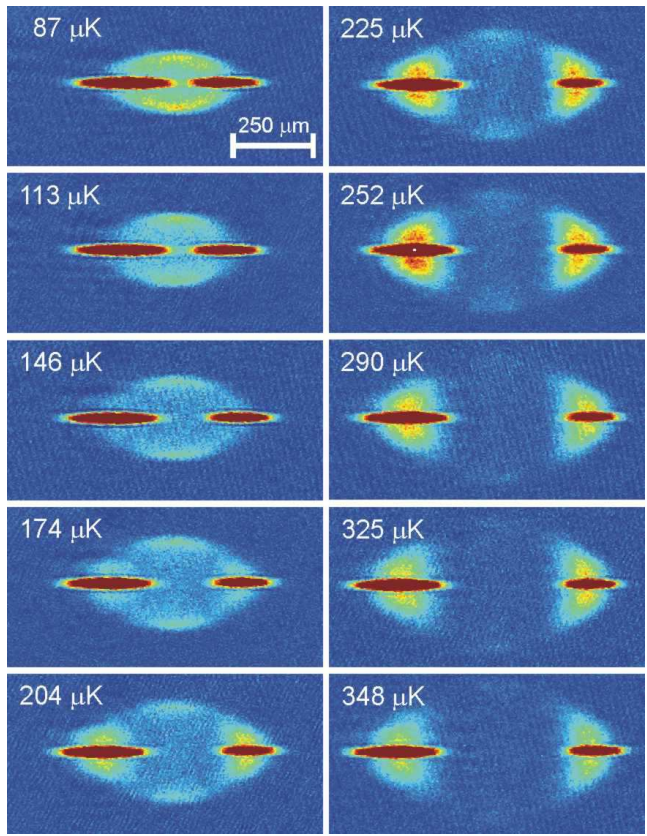}}
\caption{(Color online) Absorption images acquired at a quarter of
a radial trap period after the collision of two doubly spin
polarized Rb clouds (visible as dark ellipses) for various
collision temperatures. The halos of scattered particles have
elliptical envelopes since they are evolving in an anisotropic
harmonic trap which is weakest in the horizontal direction
($z$-direction). At the selected time of acquisition the
scattering halos have the maximum radial excursion in the trap.}
\label{figratio}
\end{figure}
Figure~2 shows absorption images of scattering acquired at
collision temperatures in the range from 87~$\mu$K to 348~$\mu$K.
Scattering halos of particles with an elliptical envelope are
clearly visible as are the outgoing clouds of unscattered atoms.
The major and minor semiaxes of the former, and the distance
between the latter, increase linearly with $\sqrt{T}$ due to the
fixed time of acquisition after collision. The total number of
scattered particles $N_{\rm sc}$ was determined by integrating the
column density over the image frame and using a suitable
interpolation to bridge the areas hidden by the outgoing clouds of
unscattered atoms.

The observed scattering yield is interpreted in terms of a
coupled-channels theoretical model that includes the ground state
singlet and triplet potentials and all spin-dependent
interactions. The triplet potential has a van der Waals $C_6$
constant of 4707 atomic units (1~atomic unit =
9.5734$\times$10$^{-26}$~J~nm$^6$) and a scattering length of
98.96 atomic units (1 atomic unit = 0.052918~nm) \cite{Marte02}.
Figure 3(a) presents the partial wave phase shifts for the $l = 0$
and 2 channels with total projection quantum number $M=4$ when two
$|F=2, M=2\rangle$ atoms collide in a total magnetic field of 0.22
mT, the bias field of this experiment (there is negligible
difference at zero field). Using Eq.~(1) these phase shifts give
the $s$-wave, $d$-wave and total cross sections shown in
Fig.~3(b). In Fig.~3(c) we present the fraction of scattered atoms
$N_{\rm sc}/N_{\rm tot}$ versus the collision temperature as
measured in our experiments. Since $N_{\rm sc}/N_{\rm tot}>20\%$
for all $T$ (i.e., large depletion), the number of scattered
particles is not proportional to the total elastic cross section
$\sigma(T)$. As a result, the observed $d$-wave resonance peak is
not very pronounced even though the total cross section grows by a
factor of $\sim 4$ with respect to the zero temperature limit.
However, when the effect of depletion is included \cite{note2} we
obtain good agreement between the experimental and theoretical
scattering fractions [Fig.~3(c)].

As is obvious from Fig.~2, the scattered particles are emitted in
spatial patterns which depend on the collision temperature. It is
possible to relate these patterns to the differential cross
section when the effects on the particle distribution of the
harmonic potential and the projection onto the imaging plane are
accounted for. As a result of the scattered particles expanding in
an anisotropic harmonic potential, the projected halos seen in
Fig.~2 have elliptical envelopes rather than the circle expected
for a free-space Newton sphere as shown in Fig.~1. However, due to
the cylindrical symmetry about the collision axis (which is
perpendicular to the optical axis of our imaging system), full 3D
tomographical information on the scattering can be extracted from
the 2D absorption images via the inverse Abel transform
\cite{Bracewell00,Dribinski02}. Applying Abel inversion to the
absorption images gives us the angular particle distribution in
the trap at the time of image acquisition, to which the
distribution at the time of collision (the free space
distribution) is related in a straightforward manner \cite{note3}.
\begin{figure}[t!]
\centerline{\includegraphics*[width=\columnwidth]{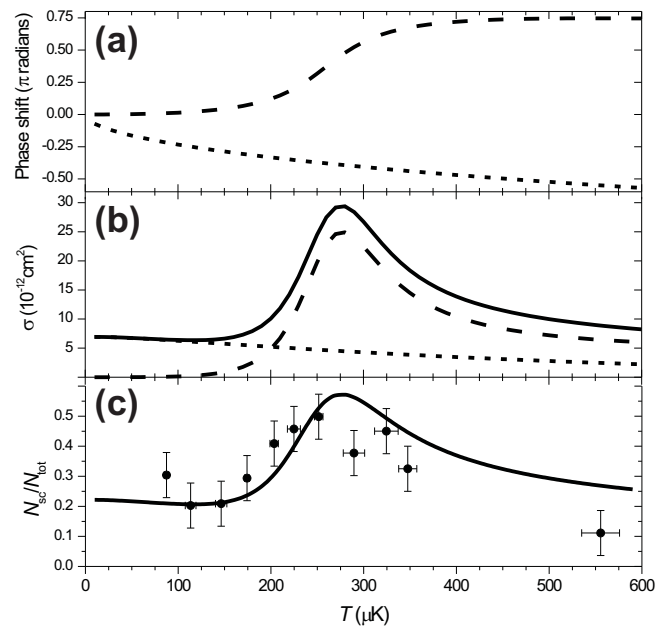}}
\caption{Dependence on collision temperature. (a) The $s$ (dotted
line) and $d$ (dashed line) partial wave phase shifts from a
simple theoretical model. (b) The $s$-wave (dotted line), $d$-wave
(dashed line) and total (solid line) cross sections calculated
from the model partial wave phase shifts. (c) The measured
scattered fraction of atoms $N_{\rm sc}/N_{\rm tot}$ (filled
circles). The black curve shows the fraction as given by the model
cross section when depletion of the colliding atom clouds is
accounted for.} \label{fig3}
\end{figure}

In Fig.~4(a) we show polar plots of the probability density
$n_{\rm sc}(\theta,T)\propto d\sigma /d\Omega$ for a scattered
particle to be emitted at the polar angle $\theta$ as determined
from the absorption images in Fig.~2. The angular distributions
for different temperatures have been normalized with respect to
each other such that $\int n_{\rm sc}(\theta,T)d\Omega=1$ for all
$T$, and were determined from the Abel inverted images by counting
the particles within angular bins at a unit sphere transformed to
the quarter period ellipsoid via the relation in
Ref.~\cite{note3}. For comparison we present in Fig.~4(b) the
temperature development of the normalized differential cross
section as predicted by Eq.~(1) using the partial wave shifts from
the previously described model. The scattering patterns of
Fig.~4(a) and Fig.~4(b) show the same behavior and the minor
discrepancies between the experimental and theoretical
distributions may be attributed to broadening effects from the
finite sizes of the colliding clouds and a small departure from an
ideal scattering geometry, both of which are not included in our
analysis method \cite{note4}. For low temperatures the scattering
is $s$-wave dominated and isotropic. However, at the onset of the
$d$-wave scattering resonance the $s$ and $d$ partial wave
amplitudes interfere constructively in the radial direction and
destructively in the axial direction. Above the $d$-wave resonance
the scattering pattern is $d$-wave dominated, but non-vanishing
$s$-wave scattering gives rise to destructive interference in the
radial direction and constructive interference in the axial
direction. It is possible to evoke analogies to Young's well-known
double slit experiment. In the present experiment the $s$ and $d$
state each act as a slit and the distributions in Fig.~4 are
effectively fringe patterns resulting from the absence of
which-way-information.
\begin{figure}[t!]
\centerline{\includegraphics*[width=\columnwidth]{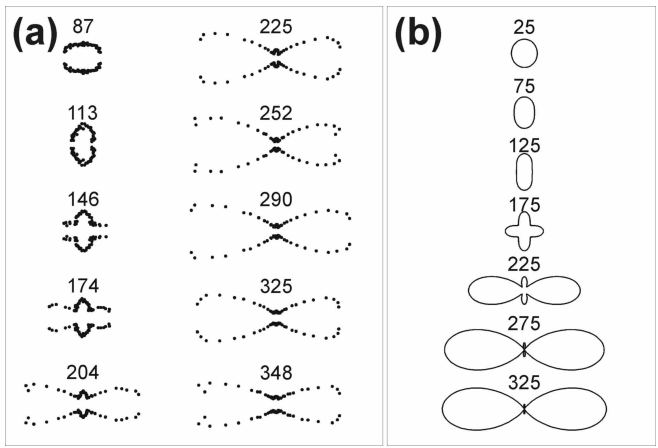}}
\caption{Polar plots of the normalized angular scattering
probability density for different collision temperatures in
$\mu$K. (a) Experimental results from the absorption images of
Fig.~2 after Abel inversion and a transformation from trap to free
space. (b) Characteristic patterns as predicted by Eq.~(1) using
the partial wave shifts from our theoretical model.}
\label{fig4}
\end{figure}

In conclusion, we have reported direct imaging of the scattered
atoms in cold collisions of doubly spin-polarized $\rm ^{87}Rb$.
The emission patterns and the measured number of scattered atoms
as a function of collision energy are described well by a simple
theoretical model. The present experiment demonstrates in
particular the quantum mechanical nature of the scattering of
atoms. The underlying quantum mechanics reveals itself strikingly
through the appearance of one of its most prominent features
--- interference --- and as only two states are involved in the
scattering, the interpretation becomes particularly simple. On a
more subtle level the extended version of Pauli's exclusion
principle gives rise to the absence of odd partial waves since the
scattering particles are identical bosons. Finally, we note the
possibility of extending our method to other important low-lying
resonances \cite{Boesten1996,demarco1999}, atoms in different spin
states, and to heteronuclear collisions.

This work was supported by the Marsden Fund of New Zealand,
Contract~02UOO080. N.~K. acknowledges additional support from the
Danish National Science Research Council. We thank V. Dribinski
for providing us with a software implementation of the Abel
inversion method of ref.~\cite{Dribinski02}.


\begin{thebibliography}{10}
\bibitem{note1}
$l=\{1,3,5...\}$ terms are strictly forbidden and $l=\{4,6,8...\}$
terms ($g$-waves and higher order) have been calculated to be
completely negligible for the energy interval of interest.
\bibitem{note2}
During collision the densities $n_i$ of the clouds labeled by
$i=1,2$ are governed by
  the two coupled nonlinear partial differential equations $\partial_t
  n_i(x,y,z,t)=v_i\partial_z
  n_i(x,y,z,t)-|v|\sigma({T})n_1(x,y,z,t)n_2(x,y,z,t)$. Assuming the clouds to
  have initially Gaussian density distributions in all directions, this system
  was solved numerically to calculate the fraction of scattered particles and
  assess the scaling with the coupling parameter $\sigma(T)$. Due
  to the prolate cloud shapes we neglect multiple scattering
  events.
\bibitem{note3}
A particle emitted in the polar angle $\theta_0$ at time $t=0$
from the center
  of the trap follows a trajectory so that $\tan \theta(t)= \tan \theta_0
  \omega_z \sin(\omega_r t)/\omega_r \sin(\omega_z t)$.
\bibitem{note4}The broadening implies that some of the scattered atoms will reach
their maximum radial excursion point before the time of image
aquisition. This effectively transfers them to angular bins closer
to the collision axis. The slight left-right pattern asymmetry
observable for some collision temperatures is a combined result of
the clouds having slightly different speeds in the observation
frame and a small offset on the $z$-axis between collision and
trap centers.
\end{thebibliography}
\end{document}